\documentclass[fleqn,10pt]{wlscirep}
\title{On the possibility of many-body localization in a doped Mott insulator}

\author[1,*]{Rong-Qiang He}
\author[1]{Zheng-Yu Weng}
\affil[1]{Institute for Advanced Study, Tsinghua University, Beijing 100084, China}
\affil[*]{rqhe@ruc.edu.cn}

\begin{abstract}
Many-body localization (MBL) is currently a hot issue of interacting systems, in which quantum mechanics overcomes thermalization of statistical mechanics. Like Anderson localization of non-interacting electrons, disorders are usually crucial in engineering the quantum interference in MBL. For translation invariant systems, however, the breakdown of eigenstate thermalization hypothesis due to a \emph{pure} many-body quantum effect is still unclear. Here we demonstrate a possible MBL phenomenon without disorder, which emerges in a lightly doped Hubbard model with very strong interaction. By means of density matrix renormalization group numerical calculation on a two-leg ladder, we show that whereas a single hole can induce a very heavy Nagaoka polaron, two or more holes will form bound pair/droplets which are all localized excitations with flat bands at low energy densities. Consequently, MBL eigenstates of finite energy density can be constructed as composed of these localized droplets spatially separated. We further identify the underlying mechanism for this MBL as due to a novel `Berry phase' of the doped Mott insulator, and show that by turning off this Berry phase either by increasing the anisotropy of the model or by hand, an eigenstate transition from the MBL to a conventional quasiparticle phase can be realized.
\end{abstract}
\begin{document}

\flushbottom
\maketitle
\thispagestyle{empty}


\section*{Introduction}

As one of the most intriguing quantum phenomena, Anderson localization \cite{Anderson58} predicts that noninteracting particles may become localized in a disordered media by quantum interference. For interacting particles, the survival of Anderson localization, which is now known as many-body localization (MBL), has been demonstrated by theories \cite{Basko06,Gornyi05,Imbrie14} and numerical studies \cite{Oganesyan07,Znidaric08,Pal10,Cecile10,Canovi11,Kjall14,Luitz15,Tang15}. A profound impact of MBL is that it prevents the thermalization of subsystems in
violating the eigenstate thermalization hypothesis (ETH) \cite{Deutsch91,Srednicki94,Tasaki98,Rigol07,Rigol08}, which represents a breakdown of quantum statistical mechanics. The MBL thus gives rise to the possibility to realize phenomena forbidden by quantum statistical mechanics, and to provide the protection for quantum and topological orders at finite temperatures \cite{Wootton11,Stark11,Huse13,Bela13,Bahri13}.

An important question is wether disorder is indispensable to MBL. Specifically, can MBL-like physics appear in a translation invariant system? This question has recently attracted a lot of effort and debate\cite{Grover14,Schiulaz14,Schiulaz15,Roeck14,Roeck14Scenario,Hickey14,Horssen15,Yao14,Papic15}.
A popular idea is to construct a translation invariant system involving two species of locally interacting particles with a very large mass ratio. The heavy particles move slowly and may dynamically generate an disordered potential to localize the light particles. Other ideas also involve self-generating disordering effect by, e.g., thermal fluctuations and initial conditions. However, an MBL violating ETH due to a \emph{pure} many-body quantum effect has not been addressed so far, to our best knowledge.

In this work, we show the possibility that an MBL can emerge in a standard Hubbard model for electrons on a square lattice two-leg ladder, which can be studied efficiently by density matrix renormalization group (DMRG) numerical algorithm \cite{White92}. Such a system is a translation invariant Mott insulator at half-filling when the Hubbard $U$ is reasonably large. We focus on an extreme case with $U$ very large as compared to the hopping integrals, such that when a single hole is injected, it can polarize the neighboring spins to form a bound state called a `Nagaoka polaron'. Even though each Nagaoka polaron has a very large effective mass, it still propagates coherently. However, it becomes unstable once two holes are doped into the system, which form a tightly bound pair and get self-localized in real space. Three and four holes further form droplets, which are also well localized with infinite degeneracy since each can be located anywhere in an infinitely long ladder. A many-body eigenstate at a finite doping concentration can be then constructed by these multi-hole localized `islands', which is of finite energy density. As such, the charge sector of lightly doped Hubbard model with sufficiently large but finite $U$ has an emergent MBL violating ETH and with spontaneous translational symmetry breaking in finite energy-density states. Further, an eigenstate phase transition to a conventional delocalized phase has been found by increasing the anisotropy of the hopping integrals of the ladder or by turning off the so-called phase strings \cite{LZhang2014,JHo2015} as the intrinsic Berry phase of the Hubbard model.

\section*{Many-body localization in the Hubbard model}

\subsection*{Model}

The Hamiltonian of the Hubbard model reads
\begin{equation}\label{eq:hamiltonian}
\hat{H} = -\sum_{\langle ij\rangle,\sigma} t_{ij} (c_{i\sigma}^{\dagger}c_{j\sigma}+c_{j\sigma}^{\dagger}c_{i\sigma}) + U\sum_{i}n_{i\uparrow}n_{i\downarrow},
\end{equation}
where $c_{i\sigma}^{\dagger}$ and $c_{i\sigma}$ are respectively the creation and annihilation operators of the electron on lattice site $i$ with spin $\sigma$ ($ =\:\uparrow, \downarrow$), and $n_{i\sigma} = c_{i\sigma}^{\dagger} c_{i\sigma}$. And $t_{ij}$ is the hopping integral with $\langle ij\rangle$ denoting a pair of nearest-neighbor sites $i$ and $j$, while $U> 0$ is the strength of the on-site repulsion.

\begin{figure}[h]
  \centering
  \includegraphics[width=8cm]{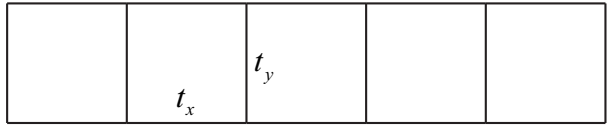}
  \caption{A two-leg Hubbard ladder with a length $L$ $(= 6)$. The hopping amplitude along the legs (rungs) is denoted by $t_x$ ($t_y$), characterized by the anisotropic parameter $\beta \equiv t_x / t_y$ with $t_x = 1$. A strong Hubbard potential of $U / t_x = 256$ will be taken in this work.}
  \label{fig:ladder}
\end{figure}

In the present work, we focus on a special case of the two-leg ladder, in which $t_{ij} \equiv t_x$ and $t_y$, respectively, along the ladder legs and rungs as illustrated in Fig. \ref{fig:ladder}. In particular, we consider a strong coupling case with $U / t_x = 256$ and use an open boundary condition in the DMRG calculation. For convenience, $t_x = 1$ is taken as the unit of energy with $\beta \equiv t_x / t_y$ describing the anisotropy of the ladder.

At half-filling, the system is a Mott insulator at a large $U$, and the ground state is spin singlet ($S=0$) with a spin excitation gap for any finite anisotropic parameter $\beta$. What we shall study is the behavior of holes doped into the Mott insulator. In the following, we investigate the single hole doping first.

\subsection*{A single hole as a Nagaoka polaron}

\begin{figure}[h]
  \centering
  \includegraphics[width=8cm]{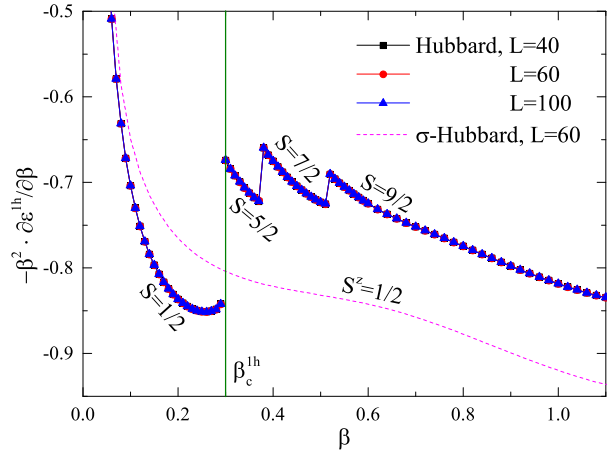}
  \caption{\label{fig:phasedigram} (Color online) The phase diagram of the single-hole ground state can be determined by the first derivative of the hole energy $\varepsilon^{\rm 1h}$ as a function of the anisotropic parameter $\beta$, which indicates a series of first-order eigenstate phase transitions characterized by different total spins $S$. At $\beta < \beta_c^{\rm 1h} = 0.304$, the single hole carries an $S=1/2$ and forms a conventional quasiparticle. For $\beta > \beta_c^{\rm 1h} $, the hole polarizes its neighboring spins to form a high-spin bound state called `Nagaoka polaron' with $S>1/2$ (cf. Fig.~\ref{fig:nhsz}). The dashed smooth curve indicates that a quasiparticle behavior is recovered in the whole range of $\beta$ of the so-called $\sigma$-Hubbard model, in which the phase string effect is turned off (see text) \cite{LZhang2014,JHo2015}.}
\end{figure}

For a doped semiconductor, a doped hole is expected to be a conventional quasiparticle satisfying the Bloch theorem, characterized by charge $+e$, spin $S=1/2$ and a well-defined momentum with a renormalized mass.  When a single hole is injected into the half-filled two-leg ladder system, which is also gapped, the new ground state in general may have a total spin $S >1/2$ due to many-body effect. Figure \ref{fig:phasedigram} shows the phase diagram determined by the first derivative of the hole energy $\varepsilon^{\rm 1h}$ over the anisotropy $\beta$ calculated by DMRG. Here $\varepsilon^{\rm 1h} \equiv E_0^{\rm 1h} - E_0^{\rm 0h}$, where $E_0^{\rm 0h}$ ($E_0^{\rm 1h}$) is the ground state energy at half-filling (one-hole-doping). Note that $E_0^{\rm 0h}$ is a smooth function of $\beta$.

As a function of $\beta$, a series of first-order jumps in $\partial \varepsilon^{\rm 1h}/\partial {\beta}$ are exhibited in Fig. \ref{fig:phasedigram}, where each $S$ labels the total spin of the corresponding ground state. Here, only in the strong rung case of $\beta < \beta_c^{\rm 1h} = 0.304$, does the hole (holon) bind with a spin-$1/2$ to form a conventional quasiparticle moving on the spin-singlet background (see below). At $\beta > \beta_c^{\rm 1h} $, the total spin of the ground state jumps from $1/2$ to $5/2$, then to $7/2$, and further to $9/2$, ...,  through a sequence of first-order eigenstate phase transitions signified by the discontinuities of $\partial \varepsilon^{\rm 1h}/\partial {\beta}$. (The total spin $S$ of a ground state is identified by the condition $E_0 = E_0(S^z)$ for $S^z = -S, -S + 1, \dotsc, S$ and $E_0 < E_0(S^z)$ for $|S^z| > S$,
where $E_0$ is the full-Hilbert-space ground state energy and $E_0(S^z)$ is the ground state energy in the subspace labelled by the total spin $z$-component $S^z$.)

\begin{figure}[h]
  \centering
  \includegraphics[width=8cm]{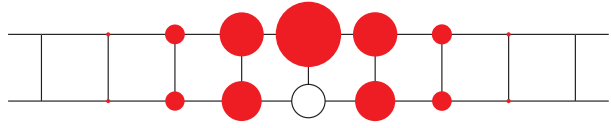}
  \caption{\label{fig:nhsz} (Color online) A typical profile of an $S=5/2$ (with $S^z=5/2$) Nagaoka polaron at $\beta = 1/3$, which is determined by the hole-spin correlation function $C_{ij}^{\rm hs}$. The radius of the red circle, indicating the magnitude of $\hat{s}^z_j$ at lattice site $j$, is proportional to the value of $C_{ij}^{\rm hs}$ with a fixed hole position at $i$ marked by the open circle. }
\end{figure}

By calculating the hole-spin correlation function $C_{ij}^{\rm hs} \equiv \langle \hat{n}_i^{\rm h} \hat{s}_j^z \rangle$, where $\hat{n}_i^{\rm h} = 1 - n_{i\uparrow} - n_{i\downarrow}$ and $\hat{s}_i^z = (n_{i\uparrow} - n_{i\downarrow}) / 2$, we find that $C_{ij}^{\rm hs}$ decays {\em exponentially} fast as the distance between lattice sites $i$ and $j$ is increased. A typical example is shown in Fig. \ref{fig:nhsz} at $\beta = 1/3$, where the spins near the hole get polarized and form a bound state with the hole, which carries a total spin $S=5/2$ and $S^z=5/2$ (totally there are $6$-fold degenerate states).
Such a hole-spin composite may be called a `Nagaoka polaron' as a reminiscence of the Nagaoka ferromagnetism \cite{Nagaoka66} induced by a single hole in the limit of $U = \infty$. Recently a doped hole as a Nagaoka polaron has been also found in a one-dimensional extended $t$-$J$ model \cite{Sano15}.

\begin{figure}[h]
  \centering
  \includegraphics[width=8cm]{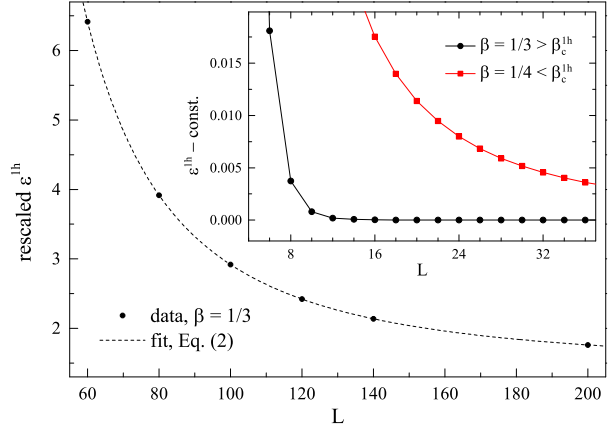}
  \caption{\label{fig:mass} (Color online) Main panel: the Nagaoka polaron energy $\varepsilon^{\rm 1h}$ with total spin $S = 5/2$ at $\beta = 1/3 > \beta_c^{\rm 1h}$. The fitting of Eq. (\ref{eq:polaronenergy}) shows that $t_{\mathrm {eff}}=5.82 \times 10^{-5}t_x$ and $L_0 = 11.97$, indicating a substantial mass enhancement $m^* /m = (t_{\mathrm {eff}} / t_x)^{-1} = 1.718 \times 10^4$. Inset:  the comparison of $\varepsilon^{\rm 1h} $ at  $\beta = 1/3$ and $1/4$, respectively. The hole forms a conventional quasiparticle (with $S=1/2$) at $\beta = 1/4< \beta_c^{\rm 1h}$, with $t_{\mathrm {eff}} / t_x=0.489$ or $m^* /m  \simeq 2.04$.  }
\end{figure}

Here we can further quantify the motion of the Nagaoka polaron by analyzing the hole energy $\varepsilon^{\rm 1h}$. As shown in Fig. \ref{fig:mass}, $\varepsilon^{\rm 1h}$ can be excellently fitted by
\begin{equation}\label{eq:polaronenergy}
\varepsilon^{\rm 1h} \approx {\rm const.} + \pi^2 t_{\mathrm {eff}} (L - L_0)^{-2},
\end{equation}
where $t_{\mathrm {eff}} $ parameterizes a renormalized hopping integral for the Nagaoka polaron along the leg (with the bare hopping integral $t_x=1$) and $L_0$ accounts for the fact that the Nagaoka polaron has a finite size. It is instructive to compare Eq.~(\ref{eq:polaronenergy}) with the ground state energy of a noninteracting free particle under an open boundary condition at large sample size $L$ in a one-dimensional chain, which follows $\varepsilon^{\rm 1h} \approx {\rm const.} + \pi^2 t_x (L + 1)^{-2}$.

As shown in the inset of Fig. \ref{fig:mass}, $\varepsilon^{\rm 1h}$ at $\beta=1/4<\beta_c^{\rm 1h} $ can be fitted with $t_{\mathrm {eff}}/t_x=0.489$, where the hole carries a spin-$1/2$. By contrast, the effective $t_{\mathrm {eff}}/t_x$ gets substantially reduced at $\beta = 1/3$ where the hole as a Nagaoka polaron carries an $S=5/2$ spin polarization (cf.  Fig. \ref{fig:nhsz}). In the main panel of Fig. \ref{fig:mass}, the fitting shows that $t_{\mathrm {eff}}/t_x= 5.82 \times 10^{-5}$, which in turn means that the effective mass of the $S=5/2$ Nagaoka polaron is substantially enhanced by $\sim  10^{4}$, as compared to that $\sim 2.04$ at $\beta=1/4$ for a conventional quasiparticle with $S=1/2$. In other words, the `band' for the $S=5/2$ Nagaoka polaron becomes quite flat.

By contrast, the Nagaoka polaron phases shown in Fig. \ref{fig:phasedigram} disappear in the so-called $\sigma$-Hubbard model \cite{JHo2015}, as shown by the dashed curve, where the doped hole remains a conventional quasiparticle throughout the accessible regime of $\beta$.  This $\sigma$-Hubbard model has been previously studied in Ref. \citenum{JHo2015}, which differs from the Hubbard model by a Berry-like phase \cite{LZhang2014} (phase string) as to be discussed in Sec. II E below. Such phase string effect is associated with the charge fluctuations (holons and doublons) of the Hubbard model. Especially, the $\sigma$-Hubbard model is essentially the same as the Hubbard model at half-filling due to the suppression of the charge fluctuations in the large-$U$ limit, and therefore the nontrivial Berry-like phase or phase string only plays a critical role upon doping as studied here.


\begin{figure}[h]
  \centering
  \includegraphics[width=8cm]{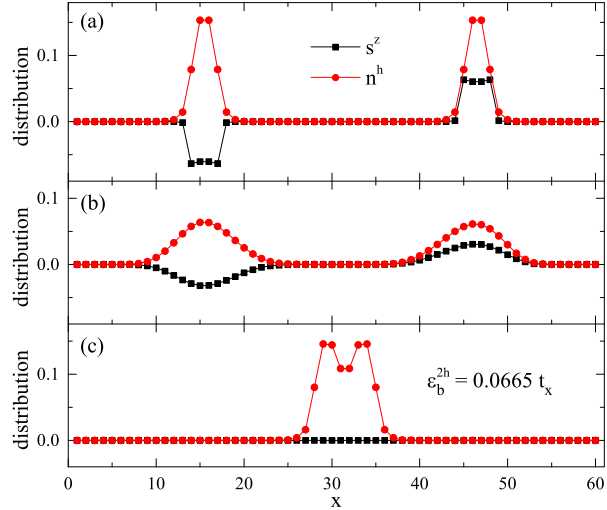}
  \caption{\label{fig:disevo} (Color online) The binding of two holes at $\beta = 2/5> \beta_c^{\rm 2h}$ ($L = 60$ and $S^z=0$) with increasing DMRG accuracy, i.e., increasing the number of states kept in DMRG from (a) to (c). Real-space distribution evolutions of spins ($s_x^z \equiv \langle \hat{s}_x^z \rangle$) and holes ($n_x^{\rm h} \equiv \langle \hat{n}_x^{\rm h} \rangle$) are shown here, where $x$ denotes the lattice sites on one leg of the ladder (the values are the same at the two sites of each rung).  From (a) to (c),  the two holes are initially located far apart and each polarizes some spins nearby to form a Nagaoka polaron, which eventually bind together to form a {\em localized} two-hole bound state fully converged within the machine precision (see text).}
\end{figure}

\subsection*{Pair localization}

Now we consider two holes doped into the half-filled system. At $\beta < \beta_c^{\rm 1h}$ where each single hole is described by a conventional spin-$1/2$ quasiparticle, no pairing between them is found, which simply remain two independent quasiparticles. However, a strong binding between the two holes occurs when the single hole becomes an $S>1/2$ Nagaoka polaron at $\beta > \beta_c^{\rm 1h}$. Note that the DMRG calculation shows the true transition point $\beta_c^{\rm 2h}=0.287\lesssim \beta_c^{\rm 1h}=0.304$.

Let us focus on an example at $\beta=2/5 >\beta_c^{\rm 2h}$, where the $S=7/2$ Nagaoka polaron becomes stable for the single hole ground state (cf. Fig. \ref{fig:phasedigram}). As illustrated in Fig.~\ref{fig:disevo}, when two holes are initially created far apart (with the total $S^z=0$), each will start to polarize its nearby spins to form a Nagaoka polaron separately with increasing the number of states kept in the DMRG simulation. Then the two wave packets of the holes spread out very slowly. Finally they meet and bind together to form a bound pair and during the process release some energy, i.e., the binding energy $\varepsilon^{\rm 2h}_{\rm b}=0.0665 $. This whole converging process in the DMRG calculation is clearly seen through the real-space distribution evolution of spins and holes, which eventually converges in Fig.~\ref{fig:disevo}(c).
\begin{figure}[h]
  \centering
  \includegraphics[width=8cm]{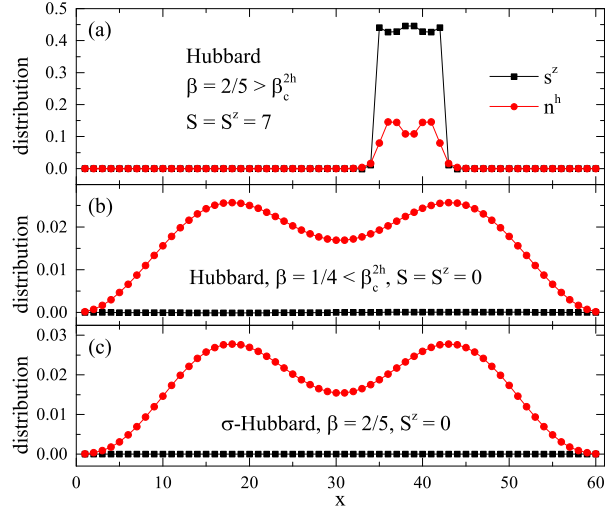}
  \caption{\label{fig:fig6} (Color online)  (a) A converged two-hole bound state localized in a different place on the ladder as compared to Fig.~\ref{fig:disevo}(c), with $S=7$ and $S^z = 7$;  (b) The collapsing of the localized bound pair profile at a smaller $\beta$ ($=1/4$); (c) The two holes are always unpaired and remain mobile after turning off the phase string (cf. Sec. II E). The two-leg ladder length $L=60$. }
\end{figure}

A further calculation verifies that the hole pair has a total spin $S=7$ in the ground state of the above case, while other $S$ states are allowed in excited states. The hole pair in Fig.~\ref{fig:disevo}(c) actually has a $15$-fold degeneracy due to the spin rotational symmetry. Furthermore, the real space location of the hole pair is also arbitrary. Indeed, another \emph{degenerate} hole pair with $S=7$ and $S^z=7$ is shown in a different position in Fig.~\ref{fig:fig6}(a). Here the ground state degeneracy is accurate up to the machine precision with the truncation error $\sim10^{-16}$ in the DMRG calculation.

The final real-space hole density profile ($n_i^{\rm h} \equiv \langle \hat{n}_i^{\rm h} \rangle$) and spin density profile ($s_i^z \equiv \langle \hat{s}_i^z \rangle$) of the hole pair are extremely `solid' [Figs. \ref{fig:disevo}(c) and \ref{fig:fig6}(a)], which do not spread out in that the first eleven digits of the $n_i^{\rm h}$ and $s_i^z$ values are unchanged when we increase the number of states kept in DMRG to reduce the truncation error down to $10^{-16}$.

However, once $\beta<\beta^{\rm 2h}_c =0.287$ or by turning off the phase string in the $\sigma$-Hubbard model \cite{JHo2015} (see Sec. II E), the pair binding and localization simultaneously collapse, as shown in Figs.~\ref{fig:fig6}(b) and \ref{fig:fig6}(c), respectively. In these cases, two holes behave like two independent quasiparticles each with spin $S=1/2$, which can be analyzed by a finite-size scaling similar to the single hole case.

\subsection*{Many-body localization}

What we have found above is that a single Nagaoka polaron is intrinsically unstable towards pairing if two holes are present. What is surprising is that, instead of being mobile, such a hole pair seems to have a vanishing effective hopping integral (or a divergent mass), as shown by the DMRG up to the machine precision. It suggests that two holes form a tightly bound object with a flat band in energy levels such that practically an infinitesimal potential will pin each pair anywhere on the ladder.

\begin{figure}[h]
  \centering
  \includegraphics[width=8cm]{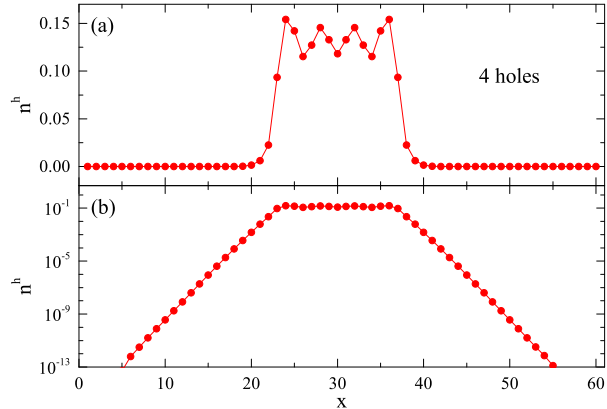}
  \caption{\label{fig:nh4h} (Color online) (a) Real-space hole distribution $n^{\rm h}$ of a 4-hole island is present at $\beta = 2/5$ and $L = 60$. (b) The logarithmic scale plot indicates an exponential fall off of the density outside the island.}
\end{figure}

\begin{figure}[h]
  \centering
  \includegraphics[width=8cm]{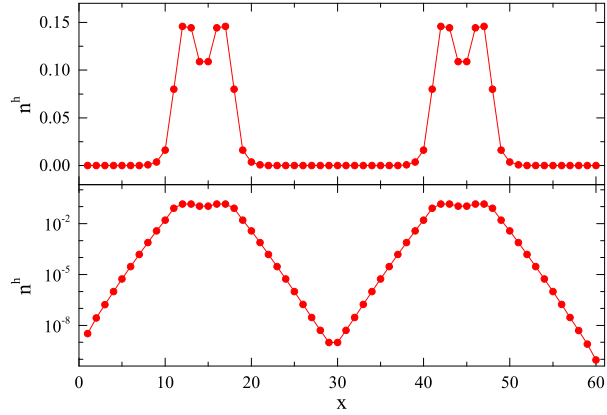}
  \caption{\label{fig:nh4h2islands} (Color online) (a) Real-space hole distribution $n^{\rm h}$ of two spatially well separated 2-hole islands of an {\em excited} state with $\beta = 2/5$ and $L = 60$. The two 2-hole islands will not further bind to form a 4-hole island in the DMRG calculation within the machine precision ($\sim 10^{-16}$). (b) The logarithmic scale plot indicates an exponential fall off of the density outside the islands. }
\end{figure}

Now let us examine what happens when more than two holes are simultaneously present in the system, which are located closely to each other. Not surprisingly, three doped holes are bound into a localized three-hole state, with a binding energy $\varepsilon^{\rm 3h}_{\rm b}=0.1408$ at $\beta=2/5$ as compared to three free Nagaoka polarons.  Four holes also form a droplet in the ground state as illustrated in Fig. \ref{fig:nh4h}. Here one may visualize first creating a pair of two-hole bound states, each of which is spatially localized as discussed above. If they are put next to each other, the holes inside them can travel in a doubled spin polarization area to lower their kinetic energy without increasing the superexchange energy of spins. Thus they will experience an attractive interaction to form a localized $4$-hole bound state, whose binding energy $\varepsilon^{\rm 4h}_{\rm b} = 0.2169 > 2 \times \varepsilon^{\rm 2h}_{\rm b} = 0.1330$ at $\beta=2/5$, accordingly. In contrast, when a pair of two-hole bound states is initially created far apart, they will localize individually and constitute an {\em excited} state, as shown in Fig.~\ref{fig:nh4h2islands}.

\begin{figure}[h]
  \centering
  \includegraphics[width=8cm]{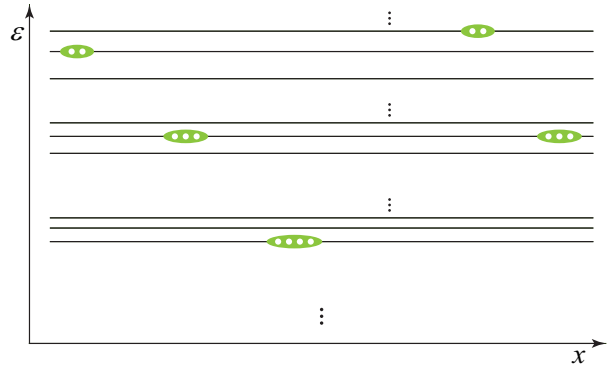}
  \caption{\label{fig:band} (Color online) Schematic illustration of finite energy density states as the flat bands filled by localized hole-islands in the charge sector. As an example, 14 holes (indicated by small white circles) are shown to form five islands (each is denoted by an ellipse). An ($n$, $m$) flat band (indicated by a long horizontal line) accommodates only the degenerate ($n$, $m$)-islands, where $n$ is the number of holes and $m$ the number of polarized spins, which never overlap in real space.}
\end{figure}

Similarly, localized $n$-hole bound states for $n > 4$ and $\varepsilon^{(n' + n'')\rm h}_{\rm b} > \varepsilon^{n' \rm h}_{\rm b} + \varepsilon^{n'' \rm h}_{\rm b}$ are generally expected in the low-energy eigenstates of the system. Each of such a localized $n$-hole bound state can be generally characterized by $n$, the number of holes, and $m$, the number of polarized spins (with a total spin $S = m/2$), which may be called an `($n$, $m$)-island'. The energy of an ($n$, $m$)-island can be denoted by $\varepsilon^{n,m}$, which is independent of where the ($n$, $m$)-island is located. Namely it lives in a `flat band'. As a result, the low-energy spectrum of the hole-doped system is a set of flat bands for the many-body charge excitations as schematically shown in Fig. \ref{fig:band}.

In this sense, the two-leg Hubbard system at large $U$ with a suitable anisotropy $\beta$, is effectively integrable for the charge sector, where an extensive set of local integrals of motion (LIOM) may be defined for the system by virtue of islands. Denote the occupation number operator of an ($n$, $m$)-island centered at position $x$ as $\hat{n}_x^{n, m}$, which takes eigenvalues 0 and 1. The corresponding effective Hamiltonian may be then written as
\begin{equation}\label{eq:heff}
\hat{H}_{\rm eff} = \sum_{x, n, m} \varepsilon^{n, m} \hat{n}_x^{n, m} + \dotsb,
\end{equation}
where the first term describes spatially well separated islands and the second term [denoted by $\dotsb$ in Eq. (\ref{eq:heff})] refers to attractive interactions between two or more islands when they are close to each other, which will lead to a new larger island with a lowered energy. In particular, in the lowest energy states (ground states), one expects all the holes grouping into a single phase-separated region as the largest island \cite{Emery90,Liu12}. But we emphasize that a many-body eigenstate as composed of smaller localized hole droplets, which are well separated along the infinite ladder, remains stable as an excited state of finite energy density according to the above numerical analysis. In other words, a vanishingly weak disorder will suffice to pin down each hole island on the ladder since its spontaneous localization has been already verified up to the machine precision in our DMRG calculation.

Therefore, $\{ \hat{n}_x^{n, m} \}$ represents an extensive set of LIOM for the charge sector of the Hubbard ladder at finite energy densities. The LIOM constrain the dynamics of the system and break ergodicity underlying quantum statistical mechanics \cite{Tasaki98,Rigol07,Rigol08}. The system prevents a complete thermalization of any given subsystems of the charge sector and thus many-body localizes\cite{Serbyn13,Kim14,Huse14,Chandran15,Ros15}. 
Evidently, the present MBL is fundamentally different from the conventional disorder-sustained MBL discussed in the literature, in which a finite strength of disorder is required.

So far the MBL in the charge sector is analyzed based on the discovery of isolated hole islands as the ground states at different hole numbers. A finite energy density state at a finite doping of holes can be constructed in terms of these localized objects of flat bands.  For all the above analysis, the effect of the spin excitations is not discussed. For example, the spins surrounding a hole island can be further excited. As the total spin increases, the size of the island may also get increased. For a sufficiently high energy density, these islands may finally overlap with each other. Whether delocalization of hole islands may occur or not, namely, if there could exist a high energy density many-body mobility edge in the many-body energy spectrum, will need further study.

Another kind of spin excitations involves that in the spin singlet background. As easily confirmed by numerical calculations, a spin triplet excitation in the background is usually delocalized with a large effective mass as the effective coupling between spins ($J = 4 t^2 / U$) is small. The exploration for the case of the simultaneous presence of hole doping and this kind of spin excitations is beyond the reach of DMRG for the current system. Without considering the above self-localization, a simple speculation for this case is that the system may be in quasi-MBL according to the theories involving heavy particles (spins here) and light particles (holes here) \cite{Grover14,Schiulaz14,Schiulaz15,Yao14,Papic15}.

\subsection*{Mechanism of the MBL: Phase string effect}

As shown in Fig. \ref{fig:phasedigram} in the single hole case, all the Nagaoka polaron states (with $S>1/2$) disappear in the so-called $\sigma$-Hubbard model \cite{JHo2015}, in which only a conventional quasiparticle phase survives (cf. the smooth dashed curve in Fig. \ref{fig:phasedigram}).
We have also seen that the localized bound pair of two holes collapses in the $\sigma$-Hubbard model [cf. Fig.~\ref{fig:fig6}(c)], where two conventional mobile quasiparticles can be identified by a finite-size scaling. Furthermore, three-hole and four-hole droplets (islands) all disappear in the $\sigma$-Hubbard model. In other words, the doped holes in the $\sigma$-Hubbard model will generally behave like conventional quasiparticles.

Similarly the doped holes in the Hubbard model can also recover a quasiparticle behavior in the strong rung limit, with $\beta$ less than the corresponding critical point, i.e., $\beta_c^{\rm 1h}$,  $\beta_c^{\rm 2h}$, ..., in the many-body eigenstates of the system with doping one hole, two holes, etc.

In the following, we point out that such transitions from the MBL to the delocalized states of conventional quasiparticles are all due to the fact that the phase string effect hidden in the Hubbard model gets either turned off or effectively `screened' under different circumstances.

Specifically, the phase string effect in the bipartite Hubbard model is manifested in the partition function as follows:\cite{LZhang2014} At any temperature and doping,
\begin{equation}\label{z}
{\cal Z}_{\mathrm {Hubb}}=\sum_{c}\tau_c^{\rm ps}\tau_c^{\rm ex} {\cal W}(c)~,
\end{equation}
where the weight $ {\cal W}(c)\geq 0$ for any closed path $c$ of displacements of holons (empty sites), doublons (doubly occupied sites) and spins (singly occupied sites). Here the phase string sign structure is Berry-phase-like as given by \cite{LZhang2014}
\begin{equation}\label{sign}
\tau_c^{\rm ps}\equiv (-1)^{N_{\downarrow}^{\rm h}(c)+N_{\downarrow}^{\rm d}(c)}
\end{equation}
which is determined by counting the total numbers of exchanges between the down spins and the holons,  $N_{\downarrow}^{\rm h}(c)$, and doublons, $N_{\downarrow}^{\rm d}(c)$, respectively,  along the closed path $c$. On the other hand,
$\tau_c^{\rm ex}\equiv (-1)^{N^{\rm h}_{\rm ex}(c)+N^{\rm d}_{\rm ex}(c)}$ is the Fermi sign structure
involving the usual exchange numbers between the holons, $N^{\rm h}_{\rm ex}(c)$, and between the doublons, $N^{\rm d}_{\rm ex}(c)$, respectively.
By contrast, in the $\sigma $-Hubbard model \cite{JHo2015}, the partition function simply reduces to
\begin{equation}
{\cal Z}_{\sigma{\rm -Hubb}}=\sum_{c}\tau_c^{\rm ex} {\cal W}(c)
\end{equation}
with $\tau_c^{\rm ps}\equiv 1$. Here the $\sigma $-Hubbard model is realized by replacing $c_{i\sigma}^{\dagger}c_{j\sigma}$ in the hopping term of the Hubbard model with
\begin{eqnarray}\label{sigmahubbard}
(1-n_{i\bar{\sigma}})c_{i\sigma}^{\dagger}c_{j\sigma} n_{j\bar{\sigma}} + n_{i\bar{\sigma}}c_{i\sigma}^{\dagger}c_{j\sigma}(1-n_{j\bar{\sigma}}) + \nonumber \\
\sigma(1-n_{i\bar{\sigma}})c_{i\sigma}^{\dagger}c_{j\sigma}(1-n_{j\bar{\sigma}}) + \sigma n_{i\bar{\sigma}}c_{i\sigma}^{\dagger}c_{j\sigma}n_{j\bar{\sigma}}
\end{eqnarray}
(where $\sigma = \pm1$, $\bar{\sigma} = -\sigma$). It differs from the Hubbard model only by inserting a $\sigma$ factor in the last two terms, which makes the phase string sign factor $\tau_c^{\rm ps}$ being eliminated in the partition function (6). \cite{LZhang2014} The two terms with $\sigma$ prefactors in (\ref{sigmahubbard}) describe exactly the exchanges between down spins and holons or doublons, respectively.
Therefore, the phase string $\tau_c^{\rm ps}$ solely captures the distinction between the Hubbard and $\sigma $-Hubbard models.

According to the above discussions, one can naturally conclude that $\tau_c^{\rm ps}$ plays the unique role responsible for the Nagaoka polaron in the one-hole-doping case, and the MBL in finite-hole-doping cases.
That is, in the Nagaoka polaron regime, a single hole has to polarize its surrounding spins to eliminate the frustration of phase string to gain kinetic energy. Since the effective mass of a Nagaoka polaron is already remarkably enhanced (e.g., $\sim10^4$ at $\beta = 1/3$), two or more of them will have strong tendency to bind and form a localized profile, which cost negligible kinetic energy of the Nagaoka polarons but gain important hopping energies for the individual holes within an enlarged profile of the new bound state. Similarly,  $\tau_c^{\rm ps}$ can also get effectively cancelled out at strong rung limit due to a tight binding of the holon with a spin-$1/2$ spinon to form a quasiparticle \cite{JHo2015}.

Finally, we note that in the large-$U$ limit, the doublon excitations are actually suppressed, while the holons appear only upon hole doping. Then the sign structure in Eq.~(\ref{sign}) reduces to that of the $t$-$J$ model \cite{WWZ2008} with $N_{\downarrow}^{\rm d}(c)\rightarrow 0$. Previously the two-leg $t$-$J$ ladder has been studied \cite{ZZ2013,ZZ2014} by DMRG at $t/J=3$ with $\beta=1$, which corresponds to $U/t=12$ in the present case (noting $J=4t^2/U$). In this regime, the phase string sign structure of Eq.~(\ref{sign}) is also shown to play a critical role \cite{ZZ2014qp,ZZ2014cm,ZSW2016}, but the behaviors of the single hole and the pairing of two holes are dramatically different from the present extremely large-$U$ case. How the two qualitative distinct pictures evolve into each other as a function of $U$ will be an interesting subject to be explored elsewhere.

\section*{Conclusion}

The two-leg large-$U$ Hubbard ladder is a spin gapped Mott insulator at half-filling, regardless of the anisotropic parameter $\beta$. We have shown in this work a possibility that a low density of holes doped into such a spin gapped (`spin liquid') system may lead to MBL at low energy densities for a sufficiently large $U$ (e.g., $U / t_x = 256$ as considered here). Although the underlying Hamiltonian is translation invariant, the eigenstates composed of localized hole-islands break translation invariance spontaneously and have low but finite energy densities, whose entanglement entropies follow an area law as the spin background, which isolates the hole islands, is short-range entangled. These excited states violate the ETH as they are subject to an extensive set of local integrals of motion.

Here it is the Berry phase of the doped Mott insulator, i.e., the phase string effect, that is responsible for the MBL. [Physically, to understand such a problem in the Born-Oppenheimer approximation, one notes that the system is essentially composed of heavy electrons (spins) and light charges (doped holes). Here the interaction between the heavy and light particles involves not only the so-called no double occupancy constraint at large $U/t$, which is usually emphasized in the literature, but also a singular Berry phase that the light particles experience, which is often omitted in the literature. The present MBL is \emph{purely} originated from the novel many-body quantum interference effect of such a Berry-like phase, with the spin background lying in a pure quantum state without thermalization, which is thus fundamentally different from the possible MBL proposed and discussed in the literature \cite{Grover14,Yao14,Papic15}.] The flat bands of charge droplets are self-trapping of holes by polarizing the surrounding spins via phase strings. In contrast, once the phase string is artificially turned off in the so-called $\sigma$-Hubbard model, the whole MBL phenomenon is gone. The delocalized quasiparticle state can be also recovered in the strong rung limit of the Hubbard model via an eigenstate transition, where the phase string effect gets effectively `screened' once a holon is tightly bound with an $S=1/2$ spinon.

The Nagaoka polarons and the MBL are also expected to emerge in higher dimensions because the phase string effect, as the underlying principle, exists irrespective of spatial dimensions \cite{WWZ2008,LZhang2014}.
The models should be also gapped at half-filling. Thus, the checkerboard Hubbard model \cite{Yao10} and the Hubbard model on the $1/5$-depleted square lattice \cite{Wu15} in two dimensions are good candidates.
We also note that the Hubbard $U$ is tunable for ultracold fermionic atoms in an optical lattice \cite{Schneider12,Jordens08,Schneider08,Bloch08}, which provides a good experimental setting to observe the Nagaoka polaron and MBL studied here.

\bibliography{ref}

\begin{thebibliography}{10}
\expandafter\ifx\csname url\endcsname\relax
  \def\url#1{\texttt{#1}}\fi
\expandafter\ifx\csname urlprefix\endcsname\relax\def\urlprefix{URL }\fi
\providecommand{\bibinfo}[2]{#2}
\providecommand{\eprint}[2][]{\url{#2}}

\bibitem{Anderson58}
\bibinfo{author}{Anderson, P.~W.}
\newblock \bibinfo{title}{Absence of diffusion in certain random lattices}.
\newblock \emph{\bibinfo{journal}{Phys. Rev.}} \textbf{\bibinfo{volume}{109}},
  \bibinfo{pages}{1492--1505} (\bibinfo{year}{1958}).

\bibitem{Basko06}
\bibinfo{author}{Basko, D.~M.}, \bibinfo{author}{Aleiner, I.~L.} \&
  \bibinfo{author}{Altshuler, B.~L.}
\newblock \bibinfo{title}{Metal-insulator transition in a weakly interacting
  many-electron system with localized single-particle states}.
\newblock \emph{\bibinfo{journal}{Ann. Phys.}} \textbf{\bibinfo{volume}{321}},
  \bibinfo{pages}{1126--1205} (\bibinfo{year}{2006}).

\bibitem{Gornyi05}
\bibinfo{author}{Gornyi, I.~V.}, \bibinfo{author}{Mirlin, A.~D.} \&
  \bibinfo{author}{Polyakov, D.~G.}
\newblock \bibinfo{title}{Interacting electrons in disordered wires: Anderson
  localization and low-$t$ transport}.
\newblock \emph{\bibinfo{journal}{Phys. Rev. Lett.}}
  \textbf{\bibinfo{volume}{95}}, \bibinfo{pages}{206603}
  (\bibinfo{year}{2005}).

\bibitem{Imbrie14}
\bibinfo{author}{Imbrie, J.~Z.}
\newblock \bibinfo{title}{On many-body localization for quantum spin chains}.
\newblock \emph{\bibinfo{journal}{arXiv:1403.7837}}  (\bibinfo{year}{2014}).

\bibitem{Oganesyan07}
\bibinfo{author}{Oganesyan, V.} \& \bibinfo{author}{Huse, D.~A.}
\newblock \bibinfo{title}{Localization of interacting fermions at high
  temperature}.
\newblock \emph{\bibinfo{journal}{Phys. Rev. B}} \textbf{\bibinfo{volume}{75}},
  \bibinfo{pages}{155111} (\bibinfo{year}{2007}).

\bibitem{Znidaric08}
\bibinfo{author}{\v{Z}nidari\v{c}, M.}, \bibinfo{author}{Prosen, T.} \&
  \bibinfo{author}{Prelov\v{s}ek, P.}
\newblock \bibinfo{title}{Many-body localization in the heisenberg $xxz$ magnet
  in a random field}.
\newblock \emph{\bibinfo{journal}{Phys. Rev. B}} \textbf{\bibinfo{volume}{77}},
  \bibinfo{pages}{064426} (\bibinfo{year}{2008}).

\bibitem{Pal10}
\bibinfo{author}{Pal, A.} \& \bibinfo{author}{Huse, D.~A.}
\newblock \bibinfo{title}{Many-body localization phase transition}.
\newblock \emph{\bibinfo{journal}{Phys. Rev. B}} \textbf{\bibinfo{volume}{82}},
  \bibinfo{pages}{174411} (\bibinfo{year}{2010}).

\bibitem{Cecile10}
\bibinfo{author}{Monthus, C.} \& \bibinfo{author}{Garel, T.}
\newblock \bibinfo{title}{Many-body localization transition in a lattice model
  of interacting fermions: Statistics of renormalized hoppings in configuration
  space}.
\newblock \emph{\bibinfo{journal}{Phys. Rev. B}} \textbf{\bibinfo{volume}{81}},
  \bibinfo{pages}{134202} (\bibinfo{year}{2010}).

\bibitem{Canovi11}
\bibinfo{author}{Canovi, E.}, \bibinfo{author}{Rossini, D.},
  \bibinfo{author}{Fazio, R.}, \bibinfo{author}{Santoro, G.~E.} \&
  \bibinfo{author}{Silva, A.}
\newblock \bibinfo{title}{Quantum quenches, thermalization, and many-body
  localization}.
\newblock \emph{\bibinfo{journal}{Phys. Rev. B}} \textbf{\bibinfo{volume}{83}},
  \bibinfo{pages}{094431} (\bibinfo{year}{2011}).

\bibitem{Kjall14}
\bibinfo{author}{Kj\"all, J.~A.}, \bibinfo{author}{Bardarson, J.~H.} \&
  \bibinfo{author}{Pollmann, F.}
\newblock \bibinfo{title}{Many-body localization in a disordered quantum ising
  chain}.
\newblock \emph{\bibinfo{journal}{Phys. Rev. Lett.}}
  \textbf{\bibinfo{volume}{113}}, \bibinfo{pages}{107204}
  (\bibinfo{year}{2014}).

\bibitem{Luitz15}
\bibinfo{author}{Luitz, D.~J.}, \bibinfo{author}{Laflorencie, N.} \&
  \bibinfo{author}{Alet, F.}
\newblock \bibinfo{title}{Many-body localization edge in the random-field
  heisenberg chain}.
\newblock \emph{\bibinfo{journal}{Phys. Rev. B}} \textbf{\bibinfo{volume}{91}},
  \bibinfo{pages}{081103} (\bibinfo{year}{2015}).

\bibitem{Tang15}
\bibinfo{author}{Tang, B.}, \bibinfo{author}{Iyer, D.} \&
  \bibinfo{author}{Rigol, M.}
\newblock \bibinfo{title}{Quantum quenches and many-body localization in the
  thermodynamic limit}.
\newblock \emph{\bibinfo{journal}{Phys. Rev. B}} \textbf{\bibinfo{volume}{91}},
  \bibinfo{pages}{161109} (\bibinfo{year}{2015}).

\bibitem{Deutsch91}
\bibinfo{author}{Deutsch, J.~M.}
\newblock \bibinfo{title}{Quantum statistical mechanics in a closed system}.
\newblock \emph{\bibinfo{journal}{Phys. Rev. A}} \textbf{\bibinfo{volume}{43}},
  \bibinfo{pages}{2046--2049} (\bibinfo{year}{1991}).

\bibitem{Srednicki94}
\bibinfo{author}{Srednicki, M.}
\newblock \bibinfo{title}{Chaos and quantum thermalization}.
\newblock \emph{\bibinfo{journal}{Phys. Rev. E}} \textbf{\bibinfo{volume}{50}},
  \bibinfo{pages}{888--901} (\bibinfo{year}{1994}).

\bibitem{Tasaki98}
\bibinfo{author}{Tasaki, H.}
\newblock \bibinfo{title}{From quantum dynamics to the canonical distribution:
  General picture and a rigorous example}.
\newblock \emph{\bibinfo{journal}{Phys. Rev. Lett.}}
  \textbf{\bibinfo{volume}{80}}, \bibinfo{pages}{1373--1376}
  (\bibinfo{year}{1998}).

\bibitem{Rigol07}
\bibinfo{author}{Rigol, M.}, \bibinfo{author}{Dunjko, V.},
  \bibinfo{author}{Yurovsky, V.} \& \bibinfo{author}{Olshanii, M.}
\newblock \bibinfo{title}{Relaxation in a completely integrable many-body
  quantum system: An \textit{Ab Initio} study of the dynamics of the highly
  excited states of 1d lattice hard-core bosons}.
\newblock \emph{\bibinfo{journal}{Phys. Rev. Lett.}}
  \textbf{\bibinfo{volume}{98}}, \bibinfo{pages}{050405}
  (\bibinfo{year}{2007}).

\bibitem{Rigol08}
\bibinfo{author}{Rigol, M.}, \bibinfo{author}{Dunjko, V.} \&
  \bibinfo{author}{Olshanii, M.}
\newblock \bibinfo{title}{Thermalization and its mechanism for generic isolated
  quantum systems}.
\newblock \emph{\bibinfo{journal}{Nature}} \textbf{\bibinfo{volume}{452}},
  \bibinfo{pages}{854--858} (\bibinfo{year}{2008}).

\bibitem{Wootton11}
\bibinfo{author}{Wootton, J.~R.} \& \bibinfo{author}{Pachos, J.~K.}
\newblock \bibinfo{title}{Bringing order through disorder: Localization of
  errors in topological quantum memories}.
\newblock \emph{\bibinfo{journal}{Phys. Rev. Lett.}}
  \textbf{\bibinfo{volume}{107}}, \bibinfo{pages}{030503}
  (\bibinfo{year}{2011}).

\bibitem{Stark11}
\bibinfo{author}{Stark, C.}, \bibinfo{author}{Pollet, L.},
  \bibinfo{author}{Imamo\ifmmode~\breve{g}\else \u{g}\fi{}lu, A. m.~c.} \&
  \bibinfo{author}{Renner, R.}
\newblock \bibinfo{title}{Localization of toric code defects}.
\newblock \emph{\bibinfo{journal}{Phys. Rev. Lett.}}
  \textbf{\bibinfo{volume}{107}}, \bibinfo{pages}{030504}
  (\bibinfo{year}{2011}).

\bibitem{Huse13}
\bibinfo{author}{Huse, D.~A.}, \bibinfo{author}{Nandkishore, R.},
  \bibinfo{author}{Oganesyan, V.}, \bibinfo{author}{Pal, A.} \&
  \bibinfo{author}{Sondhi, S.~L.}
\newblock \bibinfo{title}{Localization-protected quantum order}.
\newblock \emph{\bibinfo{journal}{Phys. Rev. B}} \textbf{\bibinfo{volume}{88}},
  \bibinfo{pages}{014206} (\bibinfo{year}{2013}).

\bibitem{Bela13}
\bibinfo{author}{Bauer, B.} \& \bibinfo{author}{Nayak, C.}
\newblock \bibinfo{title}{Area laws in a many-body localized state and its
  implications for topological order}.
\newblock \emph{\bibinfo{journal}{Journal of Statistical Mechanics: Theory and
  Experiment}} \textbf{\bibinfo{volume}{2013}}, \bibinfo{pages}{P09005}
  (\bibinfo{year}{2013}).

\bibitem{Bahri13}
\bibinfo{author}{Bahri, Y.}, \bibinfo{author}{Vosk, R.},
  \bibinfo{author}{Altman, E.} \& \bibinfo{author}{Vishwanath, A.}
\newblock \bibinfo{title}{Localization and topology protected quantum coherence
  at the edge of'hot'matter}.
\newblock \emph{\bibinfo{journal}{arXiv:1307.4092}}  (\bibinfo{year}{2013}).

\bibitem{Grover14}
\bibinfo{author}{Grover, T.} \& \bibinfo{author}{Fisher, M. P.~A.}
\newblock \bibinfo{title}{Quantum disentangled liquids}.
\newblock \emph{\bibinfo{journal}{J. Stat. Mech.: Theory Exp.}}
  \textbf{\bibinfo{volume}{2014}}, \bibinfo{pages}{P10010}
  (\bibinfo{year}{2014}).

\bibitem{Schiulaz14}
\bibinfo{author}{Schiulaz, M.} \& \bibinfo{author}{M{\"u}ller, M.}
\newblock \bibinfo{title}{Ideal quantum glass transitions: Many-body
  localization without quenched disorder}.
\newblock \emph{\bibinfo{journal}{AIP Conf. Proc.}}
  \textbf{\bibinfo{volume}{1610}}, \bibinfo{pages}{11--23}
  (\bibinfo{year}{2014}).

\bibitem{Schiulaz15}
\bibinfo{author}{Schiulaz, M.}, \bibinfo{author}{Silva, A.} \&
  \bibinfo{author}{M\"uller, M.}
\newblock \bibinfo{title}{Dynamics in many-body localized quantum systems
  without disorder}.
\newblock \emph{\bibinfo{journal}{Phys. Rev. B}} \textbf{\bibinfo{volume}{91}},
  \bibinfo{pages}{184202} (\bibinfo{year}{2015}).

\bibitem{Roeck14}
\bibinfo{author}{De~Roeck, W.} \& \bibinfo{author}{Huveneers, F.}
\newblock \bibinfo{title}{Asymptotic quantum many-body localization from
  thermal disorder}.
\newblock \emph{\bibinfo{journal}{Commun. Math. Phys.}}
  \textbf{\bibinfo{volume}{332}}, \bibinfo{pages}{1017--1082}
  (\bibinfo{year}{2014}).

\bibitem{Roeck14Scenario}
\bibinfo{author}{De~Roeck, W.} \& \bibinfo{author}{Huveneers, F.}
\newblock \bibinfo{title}{Scenario for delocalization in translation-invariant
  systems}.
\newblock \emph{\bibinfo{journal}{Phys. Rev. B}} \textbf{\bibinfo{volume}{90}},
  \bibinfo{pages}{165137} (\bibinfo{year}{2014}).

\bibitem{Hickey14}
\bibinfo{author}{Hickey, J.~M.}, \bibinfo{author}{Genway, S.} \&
  \bibinfo{author}{Garrahan, J.~P.}
\newblock \bibinfo{title}{Signatures of many-body localisation in a system
  without disorder and the relation to a glass transition}.
\newblock \emph{\bibinfo{journal}{arXiv:1405.5780}}  (\bibinfo{year}{2014}).

\bibitem{Horssen15}
\bibinfo{author}{van Horssen, M.}, \bibinfo{author}{Levi, E.} \&
  \bibinfo{author}{Garrahan, J.~P.}
\newblock \bibinfo{title}{Dynamics of many-body localization in a
  translation-invariant quantum glass model}.
\newblock \emph{\bibinfo{journal}{Phys. Rev. B}} \textbf{\bibinfo{volume}{92}},
  \bibinfo{pages}{100305(R)} (\bibinfo{year}{2015}).

\bibitem{Yao14}
\bibinfo{author}{Yao, N.~Y.}, \bibinfo{author}{Laumann, C.~R.},
  \bibinfo{author}{Cirac, J.~I.}, \bibinfo{author}{Lukin, M.~D.} \&
  \bibinfo{author}{Moore, J.~E.}
\newblock \bibinfo{title}{Quasi many-body localization in translation invariant
  systems}.
\newblock \emph{\bibinfo{journal}{arXiv:1410.7407}}  (\bibinfo{year}{2014}).

\bibitem{Papic15}
\bibinfo{author}{Papic, Z.}, \bibinfo{author}{Stoudenmire, E.} \&
  \bibinfo{author}{Abanin, D.~A.}
\newblock \bibinfo{title}{Is many-body localization possible in the absence of
  disorder?}
\newblock \emph{\bibinfo{journal}{arXiv:1501.00477}}  (\bibinfo{year}{2015}).

\bibitem{White92}
\bibinfo{author}{White, S.~R.}
\newblock \bibinfo{title}{Density matrix formulation for quantum
  renormalization groups}.
\newblock \emph{\bibinfo{journal}{Phys. Rev. Lett.}}
  \textbf{\bibinfo{volume}{69}}, \bibinfo{pages}{2863--2866}
  (\bibinfo{year}{1992}).

\bibitem{LZhang2014}
\bibinfo{author}{Zhang, L.} \& \bibinfo{author}{Weng, Z.-Y.}
\newblock \bibinfo{title}{Sign structure, electron fractionalization, and
  emergent gauge description of the hubbard model}.
\newblock \emph{\bibinfo{journal}{Phys. Rev. B}} \textbf{\bibinfo{volume}{90}},
  \bibinfo{pages}{165120} (\bibinfo{year}{2014}).

\bibitem{JHo2015}
\bibinfo{author}{Zhu, Z.}, \bibinfo{author}{Weng, Z.-Y.} \&
  \bibinfo{author}{Ho, T.-L.}
\newblock \bibinfo{title}{Spin and charge modulations in a single hole doped
  hubbard ladder - verification with optical lattice experiments}.
\newblock \emph{\bibinfo{journal}{arXiv:1510.00035}}  (\bibinfo{year}{2015}).

\bibitem{Nagaoka66}
\bibinfo{author}{Nagaoka, Y.}
\newblock \bibinfo{title}{Ferromagnetism in a narrow, almost half-filled $s$
  band}.
\newblock \emph{\bibinfo{journal}{Phys. Rev.}} \textbf{\bibinfo{volume}{147}},
  \bibinfo{pages}{392--405} (\bibinfo{year}{1966}).

\bibitem{Sano15}
\bibinfo{author}{Sano, K.} \& \bibinfo{author}{Takano, K.}
\newblock \bibinfo{title}{Ferromagnetic clouds caused by hole motion in a
  one-dimensional $t$-$j$ model}.
\newblock \emph{\bibinfo{journal}{arXiv:1510.06881}}  (\bibinfo{year}{2015}).

\bibitem{Emery90}
\bibinfo{author}{Emery, V.~J.}, \bibinfo{author}{Kivelson, S.~A.} \&
  \bibinfo{author}{Lin, H.~Q.}
\newblock \bibinfo{title}{Phase separation in the \textit{t} - \textit{J}
  model}.
\newblock \emph{\bibinfo{journal}{Phys. Rev. Lett.}}
  \textbf{\bibinfo{volume}{64}}, \bibinfo{pages}{475--478}
  (\bibinfo{year}{1990}).

\bibitem{Liu12}
\bibinfo{author}{Liu, L.}, \bibinfo{author}{Yao, H.}, \bibinfo{author}{Berg,
  E.}, \bibinfo{author}{White, S.~R.} \& \bibinfo{author}{Kivelson, S.~A.}
\newblock \bibinfo{title}{Phases of the infinite $u$ hubbard model on square
  lattices}.
\newblock \emph{\bibinfo{journal}{Phys. Rev. Lett.}}
  \textbf{\bibinfo{volume}{108}}, \bibinfo{pages}{126406}
  (\bibinfo{year}{2012}).

\bibitem{Serbyn13}
\bibinfo{author}{Serbyn, M.}, \bibinfo{author}{Papi\'{c}, Z.} \&
  \bibinfo{author}{Abanin, D.~A.}
\newblock \bibinfo{title}{Local conservation laws and the structure of the
  many-body localized states}.
\newblock \emph{\bibinfo{journal}{Phys. Rev. Lett.}}
  \textbf{\bibinfo{volume}{111}}, \bibinfo{pages}{127201}
  (\bibinfo{year}{2013}).

\bibitem{Kim14}
\bibinfo{author}{Kim, I.~H.}, \bibinfo{author}{Chandran, A.} \&
  \bibinfo{author}{Abanin, D.~A.}
\newblock \bibinfo{title}{Local integrals of motion and the logarithmic
  lightcone in many-body localized systems}.
\newblock \emph{\bibinfo{journal}{arXiv:1412.3073}}  (\bibinfo{year}{2014}).

\bibitem{Huse14}
\bibinfo{author}{Huse, D.~A.}, \bibinfo{author}{Nandkishore, R.} \&
  \bibinfo{author}{Oganesyan, V.}
\newblock \bibinfo{title}{Phenomenology of fully many-body-localized systems}.
\newblock \emph{\bibinfo{journal}{Phys. Rev. B}} \textbf{\bibinfo{volume}{90}},
  \bibinfo{pages}{174202} (\bibinfo{year}{2014}).

\bibitem{Chandran15}
\bibinfo{author}{Chandran, A.}, \bibinfo{author}{Kim, I.~H.},
  \bibinfo{author}{Vidal, G.} \& \bibinfo{author}{Abanin, D.~A.}
\newblock \bibinfo{title}{Constructing local integrals of motion in the
  many-body localized phase}.
\newblock \emph{\bibinfo{journal}{Phys. Rev. B}} \textbf{\bibinfo{volume}{91}},
  \bibinfo{pages}{085425} (\bibinfo{year}{2015}).

\bibitem{Ros15}
\bibinfo{author}{Ros, V.}, \bibinfo{author}{M{\"u}ller, M.} \&
  \bibinfo{author}{Scardicchio, A.}
\newblock \bibinfo{title}{Integrals of motion in the many-body localized
  phase}.
\newblock \emph{\bibinfo{journal}{Nuclear Physics B}}
  \textbf{\bibinfo{volume}{891}}, \bibinfo{pages}{420--465}
  (\bibinfo{year}{2015}).

\bibitem{WWZ2008}
\bibinfo{author}{Wu, K.}, \bibinfo{author}{Weng, Z.~Y.} \&
  \bibinfo{author}{Zaanen, J.}
\newblock \bibinfo{title}{Sign structure of the $t\text{-}j$ model}.
\newblock \emph{\bibinfo{journal}{Phys. Rev. B}} \textbf{\bibinfo{volume}{77}},
  \bibinfo{pages}{155102} (\bibinfo{year}{2008}).

\bibitem{ZZ2013}
\bibinfo{author}{Zhu, Z.}, \bibinfo{author}{Jiang, H.-C.}, \bibinfo{author}{Qi,
  Y.}, \bibinfo{author}{Tian, C.-S.} \& \bibinfo{author}{Weng, Z.-Y.}
\newblock \bibinfo{title}{{Strong correlation induced charge localization in
  antiferromagnets.}}
\newblock \emph{\bibinfo{journal}{Sci. Rep.}} \textbf{\bibinfo{volume}{3}},
  \bibinfo{pages}{2586} (\bibinfo{year}{2013}).

\bibitem{ZZ2014}
\bibinfo{author}{Zhu, Z.}, \bibinfo{author}{Jiang, H.-C.},
  \bibinfo{author}{Sheng, D.~N.} \& \bibinfo{author}{Weng, Z.-Y.}
\newblock \bibinfo{title}{{Nature of strong hole pairing in doped Mott
  antiferromagnets.}}
\newblock \emph{\bibinfo{journal}{Sci. Rep.}} \textbf{\bibinfo{volume}{4}},
  \bibinfo{pages}{5419} (\bibinfo{year}{2014}).

\bibitem{ZZ2014qp}
\bibinfo{author}{Zhu, Z.} \& \bibinfo{author}{Weng, Z.-Y.}
\newblock \bibinfo{title}{{Quasiparticle collapsing in an anisotropic t-J
  ladder}}.
\newblock \emph{\bibinfo{journal}{arXiv:1409.3241}}  (\bibinfo{year}{2014}).

\bibitem{ZZ2014cm}
\bibinfo{author}{Zhu, Z.} \emph{et~al.}
\newblock \bibinfo{title}{{Charge modulation as fingerprints of phase-string
  triggered interference}}.
\newblock \emph{\bibinfo{journal}{Phys. Rev. B}} \textbf{\bibinfo{volume}{92}},
  \bibinfo{pages}{35113} (\bibinfo{year}{2015}).

\bibitem{ZSW2016}
\bibinfo{author}{Zhu, Z.}, \bibinfo{author}{Sheng, D.~N.} \&
  \bibinfo{author}{Weng, Z.-Y.}
\newblock \bibinfo{title}{Breakdown of the bloch-wave behavior for a single
  hole in a gapped antiferromagnet}.
\newblock \emph{\bibinfo{journal}{arXiv:1601.00655}}  (\bibinfo{year}{2016}).

\bibitem{Yao10}
\bibinfo{author}{Yao, H.} \& \bibinfo{author}{Kivelson, S.~A.}
\newblock \bibinfo{title}{Fragile mott insulators}.
\newblock \emph{\bibinfo{journal}{Phys. Rev. Lett.}}
  \textbf{\bibinfo{volume}{105}}, \bibinfo{pages}{166402}
  (\bibinfo{year}{2010}).

\bibitem{Wu15}
\bibinfo{author}{Wu, H.-Q.}, \bibinfo{author}{He, R.-Q.},
  \bibinfo{author}{Meng, Z.~Y.} \& \bibinfo{author}{Lu, Z.-Y.}
\newblock \bibinfo{title}{Theoretical prediction of fragile mott insulators on
  plaquette hubbard lattices}.
\newblock \emph{\bibinfo{journal}{Phys. Rev. B}} \textbf{\bibinfo{volume}{91}},
  \bibinfo{pages}{125128} (\bibinfo{year}{2015}).

\bibitem{Schneider12}
\bibinfo{author}{Schneider, U.} \emph{et~al.}
\newblock \bibinfo{title}{Fermionic transport and out-of-equilibrium dynamics
  in a homogeneous hubbard model with ultracold atoms}.
\newblock \emph{\bibinfo{journal}{Nat. Phy.}} \textbf{\bibinfo{volume}{8}},
  \bibinfo{pages}{213--218} (\bibinfo{year}{2012}).

\bibitem{Jordens08}
\bibinfo{author}{J{\"o}rdens, R.}, \bibinfo{author}{Strohmaier, N.},
  \bibinfo{author}{G{\"u}nter, K.}, \bibinfo{author}{Moritz, H.} \&
  \bibinfo{author}{Esslinger, T.}
\newblock \bibinfo{title}{A mott insulator of fermionic atoms in an optical
  lattice}.
\newblock \emph{\bibinfo{journal}{Nature}} \textbf{\bibinfo{volume}{455}},
  \bibinfo{pages}{204--207} (\bibinfo{year}{2008}).

\bibitem{Schneider08}
\bibinfo{author}{Schneider, U.} \emph{et~al.}
\newblock \bibinfo{title}{Metallic and insulating phases of repulsively
  interacting fermions in a 3d optical lattice}.
\newblock \emph{\bibinfo{journal}{Science}} \textbf{\bibinfo{volume}{322}},
  \bibinfo{pages}{1520--1525} (\bibinfo{year}{2008}).

\bibitem{Bloch08}
\bibinfo{author}{Bloch, I.}, \bibinfo{author}{Dalibard, J.} \&
  \bibinfo{author}{Zwerger, W.}
\newblock \bibinfo{title}{Many-body physics with ultracold gases}.
\newblock \emph{\bibinfo{journal}{Rev. Mod. Phys.}}
  \textbf{\bibinfo{volume}{80}}, \bibinfo{pages}{885--964}
  (\bibinfo{year}{2008}).

\end{thebibliography}

\section*{Acknowledgements}

We acknowledge stimulating discussions with B. Altshuler, H.-C. Jiang, D. N. Sheng, C.-S. Tian, and especially Z. Zhu. This work was supported by National Natural Science Foundation of China (Grant No. 11534007) and National Program for Basic Research of MOST of China (Grant No. 2015CB921000).  R.Q.H. was supported by China Postdoctoral Science Foundation (Grant No. 2015T80069). Computational resources were provided by National Supercomputer Center in Guangzhou with Tianhe-2 Supercomputer.

\section*{Author contributions}

R.Q.H. developed the code, conducted the DMRG simulations, and analyzed the data. Z.Y.W. and R.Q.H. discussed the results and wrote the manuscript.

\section*{Additional information}

\noindent \textbf{Competing financial interests:} The authors declare no competing financial interests.

\end{document}